\documentclass[a4paper,aps,pre,twocolumn,floats,floatats,floatfix]{revtex4-1}
\usepackage[latin1]{inputenc}
\usepackage[brazil]{babel}
\usepackage{graphicx}  
\usepackage{dcolumn}   
\usepackage{bm}        
\usepackage{natbib}
\usepackage{latexsym}
\usepackage{mathrsfs}
\usepackage{amssymb}
\usepackage{amsmath}
\usepackage{amscd}
\usepackage{color}
\usepackage{verbatim}
\usepackage{float}
\usepackage{lineno} 
\begin{document}
\title{Transição de fase no sistema de H\'enon-Heiles}
\author{H\'ercules A. Oliveira}
\email[E-mail:~]{haoj02@gmail.com}
\affiliation{Departamento de Matem\'atica, 
Universidade Tecnol\'ogica Federal do Paran\'a,
         84016-210 - Ponta Grossa, PR, Brasil}

\date{\today}


\begin{abstract}
O sistema de Hénon-Heiles foi proposto inicialmente para descrever o 
comportamento dinâmico de galáxias, mas tem sido amplamente aplicado em sistemas
dinâmicos por exibir riqueza de detalhes no espaço de fases. 
O formalismo e a din\^amica do sistema de H\'enon Heiles \'e investigada 
neste trabalho, visando uma abordagem qualitativa.
Através das Seções de Poincaré, observa-se o crescimento da região caótica no 
espaço de fases do sistema, quando a energia total aumenta. 
Ilhas de regularidade permanecem entorno dos pontos estáveis e aparecem 
fenômenos importantes para a dinâmica, como os sticky.
\\
{\bf Palavras-chave: } Sistemas hamiltonianos, sistema de H\'enon-Heiles, 
\'orbitas peri\'odicas.
\\
\\
The Hénon-Heiles system was originally proposed to describe the 
dynamical behavior of galaxies, but this system has been widely applied in 
dynamical systems by exhibit great details in phase space.
This work presents the formalism to describe H\'enon-Heiles system and
a qualitative approach of dynamics behavior.
The growth of chaotic region in phase space is observed by Poincaré Surface 
of Section when the total energy increases.
Island of regularity remain around stable points and relevants phenomena 
appear, such as sticky.
\end{abstract}

\maketitle

\section{Introdu\c{c}\~ao}
\label{Introduction}
\par 
Em 1964, Michel Hénon e Carl Heiles \cite{henon} investigaram a existência de 
integrais de movimento para um sistema Hamiltoniano particular, que hoje é 
conhecido como sistema de Hénon-Heiles. Este sistema havia sido proposto para
descrever o movimento de galáxias interagindo via força 
gravitacional. Hénon e Heiles mostraram que a energia total e o momento 
angular do sistema são constantes de movimento, ou seja, que estas grandezas 
permanecem inalteradas ao longo da evolução temporal do sistema.

O sistema de Hénon-Heiles tem sido amplamente estudado e revela importantes
características dos sistemas Hamiltonianos, como o comportamento misto do espaço
de fases, no qual apresenta simultaneamente ilhas de regularidade e regiões 
caóticas. Esse tipo de sistema pode favorecer o fenômeno de transporte 
\cite{viana1, viana2} e mostrar pistas importantes para o estudo de órbitas 
periódicas est\'aveis e instáveis. 

Baseados no trabalho de Hénon e Heiles, analisaremos em detalhe a dinâmica 
desse sistema. Busca-se descrever o sistema analiticamente através do 
formalismo Hamiltoniano. As características e o comportamento dinâmico do 
sistema no espaço de fases são estudados aravés das Seções de Poincaré.

Na seção \ref{sistham} é feita uma breve introdução ao formalismo Hamiltoniano.
Na seção \ref{sshh} encontra-se a descrição analítica do sistema de 
Hénon-Heiles, bem como as suas equações de movimento. Na seção \ref{caosef} 
faz-se a análise das transições de fase do sistema.
Na seção \ref{conclusoes} as conclusões do trabalho e os agradecimentos na
seção \ref{agra}. 
\section{Sistemas Hamiltonianos}
\label{sistham}
Os sistemas dinâmicos podem ser divididos em conservativos e dissipativos. Os 
sistemas conservativos mantém sua energia constante a medida que o tempo evolui
e são chamados de Hamiltonianos. Os sistemas dissipativos não conservam sua 
energia ao longo do tempo.  

Um sistema Hamiltoniano pode ser descrito, em termos das coordenadas 
generalizadas de posição $q$ e momentos $p$, por uma função escalar chamada de 
{\it Hamiltoniana}, na forma
\begin{equation}
H(q, p, t)= \sum_{i=1}^{N}{p_{i}\dot{q}_{i}-L(q, \dot{q}, t)},
\end{equation}
onde $L(q, \dot{q}, t)=T-V$ é uma função chamada de {\it Lagrangeana} 
\cite{goldstein, lemos}, dada pela
energia cinética das partículas $(T=\frac{1}{2}mv^{2})$ e pela energia 
potencial 
ao qual estão submetidas $(V)$ (potencial gravitacional, potencial de uma mola,
 etc).

Se a hamiltoniana não depender explicitamente do tempo $H(q, p)$ e o potencial
for conservativo, a hamiltoniana será igual a energia total do sistema, da 
forma
\begin{equation}
H=T+V=E.
\end{equation}  

Um sistema simples, que nos serve de exemplo é o composto por um corpo de massa
$m$ ligado a uma parede através de uma mola com constante elástica $k$, sobre 
uma superfície sem atrito, como ilustrado na figura (1).
 
A hamiltoniana deste sistema é escrita como
\begin{equation}
H(q, p)= \frac{p^{2}}{2m} + \frac{1}{2} kq^{2},
\label{hmola}
\end{equation}
onde $q$ representa a coordenada $x$ generalizada e $k$ pode ser escrito em 
termos da frequência angular $k=m\omega ^{2}$. A energia cinética do corpo está 
expresso em $T=\frac{p^{2}}{2m}=\frac{1}{2}mv^{2}$ e a energia potencial da mola
é $V=\frac{1}{2}kq^{2}$.

\begin{figure}[ht!]
\begin{center}
\includegraphics[height=2.8cm,clip]{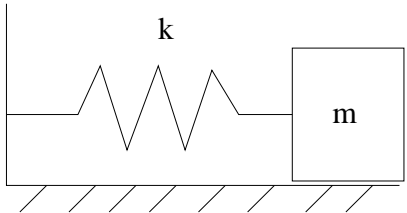}
\caption{\footnotesize{Representação de um sistema massa-mola.}}
\end{center}
\label{mola}
\end{figure}

As equações que regem o movimento dos sistemas hamiltonianos são descritas por
\cite{goldstein, lemos}
\begin{equation}
\dot{q}=\frac{\partial H}{\partial p}, \qquad
\dot{p}=-\frac{\partial H}{\partial q},
\label{eqmov}
\end{equation}
onde $\dot{q}=dq/dt$ (o ponto significa derivada temporal). Estas equações 
são as mesmas descritas pela lei de Newton onde, $\dot{q}=v$ e $\dot{p}=F$, 
correspondem à velocidade do corpo e à força a qual está submetido.

Para o sistema massa-mola as equações de movimento possuem a forma
\begin{equation}
\dot{q}=\frac{\partial H}{\partial p}=\frac{p}{m}, \qquad
\dot{p}=-\frac{\partial H}{\partial q}=-kq.
\label{eqmmola}
\end{equation}

Da equação (\ref{eqmmola}) pode-se notar que as equações de movimento resultam
no momento linear do corpo $(p=mv)$ e da força restauradora da mola 
$(F=-kq=-kx)$, o que era de se esperar.

Pode-se ainda, determinar analiticamente se um sistema é dissipativo ou 
conservativo. Adotando um sistema de dimensão $n$, podemos considerar que as $n$
variáveis  formam um volume no espaço de fases. Se esse volume permanece 
constante ou inalterado com o passar do tempo, o sistema é dito conservativo. 
Se o volume do espaço de fases contrair-se, o sistema é dissipativo.

Para verificar isso tomaremos uma superfície fechada arbitraria $S(t)$ de 
volume $V(t)$ no espaço de fases. Após um tempo infinitesimal $dt$, $S(t)$ 
evolui para $S(t+dt)$. O novo volume é dado pelo volume antigo mais a evolução
infinitesimal da superfície em todas as $n$ direções \cite{monteiro}
\begin{equation}
V(t+dt) = V(t) + \int{[(\vec{n} \cdot \vec{f})dt] dS }
\label{vefe}
\end{equation}

Multiplicando os dois lados da equação (\ref{vefe}) por $(1/dt)$, considerando 
que a evolução do volume também é infinitesimal, que a superfície é fechada 
 e que $V(t+dt) -V(t)$ é a variação do volume, temos

\begin{equation}
\frac{dV}{dt} = \oint_{S}{(\vec{n} \cdot \vec{f}) dS }.
\label{varvol}
\end{equation}

Utilizando o teorema do divergente
\begin{equation}
\oint_{S}{(\vec{n} \cdot \vec{f}) dS }=\int_{V}{(\vec{\nabla} \cdot \vec{f}) dV},
\label{tdiv}
\end{equation}

obtemos

\begin{equation}
\frac{dV}{dt} = \int_{V}{(\vec{\nabla} \cdot \vec{f}) dV}.
\label{varvol1}
\end{equation}

O sistema é conservativo quando $\frac{dV}{dt} =0$, ou seja, quando o divergente
do campo $\vec{f}$ for nulo
\begin{equation}
\vec{\nabla} \cdot \vec{f} = \frac{\partial f_{1}}{\partial q_{1}}+
 \frac{\partial f_{2}}{\partial q_{2}}+ \dots + 
\frac{\partial f_{n}}{\partial q_{n}}=0.
\end{equation}

Dessa forma a integral em (\ref{varvol1}) é nula. 
Para um sistema dissipativo $\frac{dV}{dt} < 0$ e 
$\vec{\nabla} \cdot \vec{f} < 0$. Se $\vec{\nabla} \cdot \vec{f} > 0$ ele é 
expansivo.

Para exemplicar esse conceito, pode-se escrever um sistema dinâmico na forma
\begin{equation}
\dot{\vec{x}} = \vec{f}(q_{i}, p_{i}),
\label{sd1}
\end{equation}
 onde o campo vetorial $\vec{f}(q_{i}, p_{i})$ é representado pelas equações
de movimento, no caso de sistemas Hamiltonianos. Dessa forma, o campo de 
vetores para o sistema massa-mola é descrito como
\begin{equation}
\frac{\partial f_{1}}{\partial q_{1}} = 
 \frac{\partial f_{1}}{\partial q} =  \frac{\partial \dot{q}}{\partial q} = 0,
\qquad
\frac{\partial f_{2}}{\partial q_{2}} = 
 \frac{\partial f_{2}}{\partial p} = \frac{\partial \dot{p}}{\partial p} = 0 .
\label{sd2}
\end{equation} 
Então 
\begin{equation}
\vec{\nabla} \cdot \vec{f} = \frac{\partial f_{1}}{\partial q}+
 \frac{\partial f_{2}}{\partial p} = 0.
\end{equation}
\section{O sistema de Hénon-Heiles}
\label{sshh}
O sistema generalizado de Hénon-heiles é descrito pela hamiltoniana
\begin{equation}
H(x, y, p_{x}, p_{y})=\frac{p^{2}_{x}}{2m}+\frac{p^{2}_{y}}{2m}+
\frac{1}{2}k(x^{2}+y^{2}) + \lambda (x^{2}y - \frac{1}{3}y^{3}),
\label{hamgen}
\end{equation}
onde $k$ e $\lambda$ são coeficiente constantes. Quando usamos unidades 
adimensionais $(m=1, k=1, \lambda =1)$, temos
\begin{equation}
H(x, y, p_{x}, p_{y})=\frac{p^{2}_{x}}{2}+\frac{p^{2}_{y}}{2}+
\frac{1}{2}x^{2}+\frac{1}{2}y^{2} + x^{2}y - \frac{1}{3}y^{3}.
\label{ham}
\end{equation}
A hamiltoniana (\ref{ham}) descreve a energia total de uma partícula com 
energia cinética $T=\frac{p^{2}_{x}}{2}+\frac{p^{2}_{y}}{2}$ sujeita a um potencial
\cite{lichtenberg}
\begin{equation}
V=\frac{1}{2}x^{2}+\frac{1}{2}y^{2} + x^{2}y - \frac{1}{3}y^{3}, 
\label{phen}
\end{equation}
que pode ser visto na figura (2) em forma de curvas de níveis. Nesta figura, o
eixo $z$ representa a energia do potencial para a região selecionada em $x$ e 
$y$.
As trajetórias da partícula estão contidas num plano $(xy)$ e sua 
energia total é conservada $(H=E)$, como pode ser visto nas figuras (2) e (3).
A figura (2) nos mostra uma transição de potencial de forma contínua, enquanto 
a figura (3) tem potenciais discretos, com valores específicos. 
A figura (3) mostra claramente o formato do potencial para diferentes valores
de energia potencial. Nesta figura, pode ser visto os potenciais para 
$E=1/24$, $E=1/10$, $E=1/8$ e $E=1/6$, do potencial mais interno para o mais 
externo, respectivamente.
\begin{figure}[ht!]
\begin{center}
\includegraphics[height=6.6cm,clip]{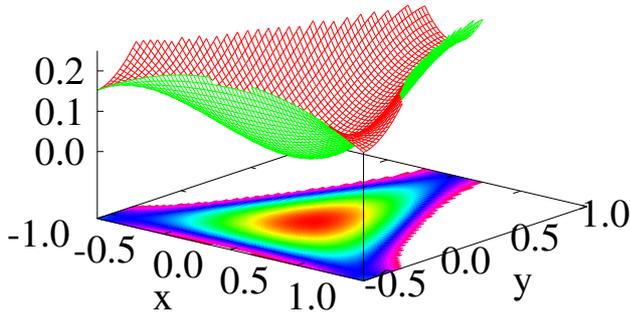}
\caption{\footnotesize{Potencial de Henón-Heiles. Plot da equação (\ref{phen})
do potencial (no eixo $z$) pelas coordenadas $x$ e $y$.}}
\end{center}
\label{pot1}
\end{figure}
\begin{figure}[ht!]
\begin{center}
\includegraphics[height=6cm,clip]{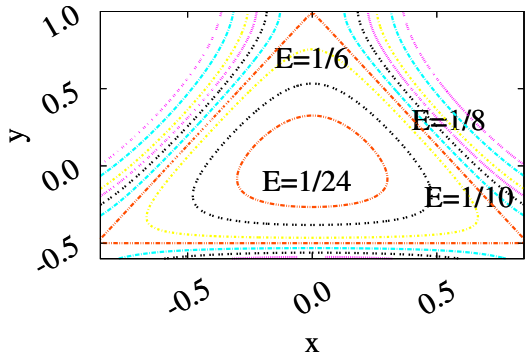}
\caption{\footnotesize{Potencial referente a equação (\ref{phen}) em duas 
dimensões com transição 
discreta, apresentando os valores de energia $E=1/24$, $E=1/10$, $E=1/8$ e 
$E=1/6$, do potencial mais interno para o mais externo, respectivamente. 
(Considere E o valor da energia potencial, ou seja, V=E apenas para esta 
figura.)}}
\end{center}
\label{pot2}
\end{figure}

Nota-se que a partícula se movimenta num plano $xy$ em formato triangular. Esse 
tipo de sistema, onde partículas são confinadas em determinadas regiões do 
espaço, também é conhecido por bilhar \cite{her}.

As equações de movimento, de (\ref{eqmov}), para o sistema de Henón-Heiles, são
\begin{equation}
\dot{x}= p_{x}, 
\label{xh}
\end{equation}
\begin{equation}
\dot{y}= p_{y},
\label{yh}
\end{equation}
\begin{equation}
\dot{p_{x}}= -x -2xy, 
\label{pxh}
\end{equation}
e
\begin{equation}
\dot{p_{y}}= -y -x^{2} +y^{2}.
\label{pyh}
\end{equation}
\\
\par As duas primeiras equações (\ref{xh}) e (\ref{yh}) são as velocidades, 
em $x$ e 
em $y$ da partícula, pois $p$ é o momento linear dado por $p=mv$. Como estamos 
usando unidades adimensionais $p=v$.
As equações (\ref{pxh}) e (\ref{pyh}) são as forças nas quais a partícula está 
submetida em $x$ e $y$, pois da segunda lei de Newton, $F=dp/dt$.

Através das equações de movimento (\ref{xh}), (\ref{yh}), (\ref{pxh}) e 
(\ref{pyh}) podemos mostrar que o sistema é conservativo, ou seja, mantém sua
energia constante no tempo. 

Fazendo
\begin{equation}
\vec{\nabla} \cdot \vec{f} = \frac{\partial f_{1}}{\partial x} +  
\frac{\partial f_{2}}{\partial y} +  \frac{\partial f_{3}}{\partial p_{x}} +
 \frac{\partial f_{4}}{\partial p_{y}} 
\end{equation}
\begin{equation}
\vec{\nabla} \cdot \vec{f} = 0+0+0+0=0.
\end{equation}

 Ou até mesmo

\begin{equation}
\vec{\nabla} \cdot \vec{f} = \frac{\partial f_{1}}{\partial p_{x}} +  
\frac{\partial f_{2}}{\partial p_{y}} +  \frac{\partial f_{3}}{\partial x} +
 \frac{\partial f_{4}}{\partial y} 
\end{equation}
\begin{equation}
\vec{\nabla} \cdot \vec{f} = 1+1-1-2y-1+2y=0.
\end{equation}

Como $\vec{\nabla} \cdot \vec{f}=0$, diz-se que o volume no espaço de fases 
do sistema
permanece constante conforme o tempo evolui. Isso nos indica que a energia 
total do sistema é a mesma ao longo do tempo, portanto, o sistema é 
conservativo.
\section{Caos no espaço de fases}
\label{caosef}
Nesta seção mostraremos como o sistema de Hénon-Heiles se comporta no espaço de 
fases. Estamos interessados em observar a transição de fase neste espaço,
de regular a caótico, através da técnica, inicialmente empregada por Henri 
Poincaré \cite{poincare}, da Seção de Poincaré. Esta técnica mostra o retrato
de fases do sistema em duas dimensões com a limitação de algumas coordenadas.

Na figura (4) podemos observar a Seção de Poincaré (retrato de 
fases) para o sistema de Hénon-Heiles. Esta figura foi construida através da 
evolução temporal do sistema com $100$ trajetórias e tempo computacional de 
3000. Através do Fortran 77 realizamos a integração numérica do sistema com o
método de Runge-Kutta de quarta ordem e gerador numérico de condições iniciais
aleatórias.
\begin{figure}[ht!]
\begin{center}
\includegraphics[height=7cm,clip]{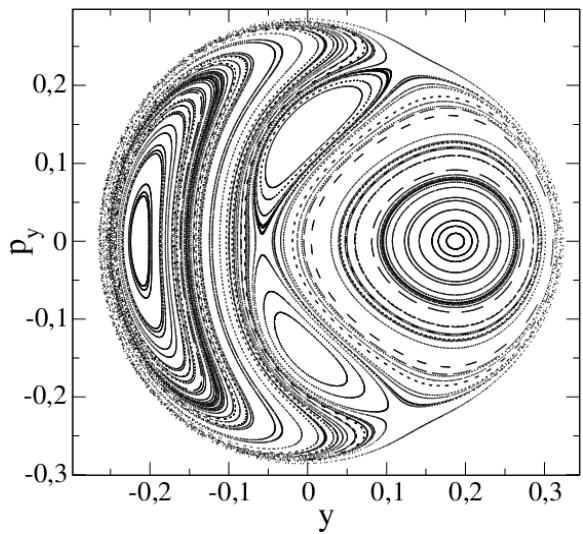}
\caption{\footnotesize{Seção de Poincaré para o sistema de Hénon-Heiles, com 
energia total igual a $1/24=0,041666667$, $100$ trajetórias e um tempo de 
evolução de $3000$.}}
\end{center}
\label{poine_124}
\end{figure}

A energia total do sistema, na figura (4), é de $E=1/24$. 
A seção de Poincaré é escrita em termos das variáveis dinâmicas $p_{y}$ 
(momento linear no eixo $y$) e
$y$ (coordenada), quando temos $x=0$ e $p_{x}>0$. O sistema pode ser 
visto como sendo constituido por uma partícula confinada num potencial 
quase triangular (ver figura 3). A seção de Poincaré é construida toda vez que 
esta partícula passa por $x=0$, com velocidade positiva em $x$, e então 
marcamos a sua posição e velocidade em $y$.

Para esta energia o sistema é dito {\it integrável} \cite{goldstein, lemos}. 
Isto significa que, num sistema Hamiltoniano conservativo com $n$ graus de 
liberdade, este é dito integrável se existem $n$ constantes de movimento 
$(F_{i})$, ou integrais de movimento como em \cite{henon}, independentes em 
involução, ou seja:
\begin{eqnarray}
\{ F_{i}, H\}=0, && \qquad i=1, \cdots , n; \nonumber \\
 \{ F_{i}, F_{j}\}=0, && \qquad i,j=1, \cdots , n,
\end{eqnarray}
onde 
\begin{equation}
\{ F_{i}, H\}=\sum_{i}^{n}{\left(\frac{\partial F_{i}}{\partial q_{i}}
\frac{\partial H}{\partial p_{i}}-
\frac{\partial F_{i}}{\partial p_{i}}
\frac{\partial H}{\partial q_{i}}\right) }
\end{equation} 
é o Parênteses de Poisson. O sistema de Hénon-Heiles possui $4$ graus de 
liberdade, e no caso integrável, o mesmo número de constantes de movimento,
identificadas, apenas duas até o momento, analiticamente como a energia total 
$(E)$, o momento angular $(\vec{l}=\vec{r}\times \vec{p})$ 
\cite{henon, goldstein}.
As outras duas integrais de movimento são combinações das coordenadas e 
momentos.

Voltando a figura (4), pode-se ver quatro aglomerados de ilhas. 
Estas ilhas circundam pontos de equilíbrio estáveis, chamados de pontos 
elípticos \cite{lichtenberg, wiggins}.
As linhas fechadas (ilhas) no retrato de fases são órbitas periódicas e 
representam a evolução de trajetórias quase-regulares. 
Este tipo de trajetória pode ser visto na figura (5) com mais detalhes. 

A regularidade de um sistema é medida pela regularidade de suas trajetórias. 
Uma trajetória regular evolui temporalmente a partir de determinadas 
coordenadas no espaço de fases e retorna a elas após um certo tempo.
Uma trajetória quase-regular procede da mesma forma, mas nunca retorna ao 
mesmo ponto, limitando-se a passar muito próxima das coordenadas iniciais. 
Uma trajetória caótica não expressa nenhuma regularidade. Esta pode nunca 
mais se aproximar ou retornar de suas coordenadas iniciais. 

A trajetória caótica pode ser vista na figura (6). 
\begin{figure}[ht!]
\begin{center}
\includegraphics[height=7cm,clip]{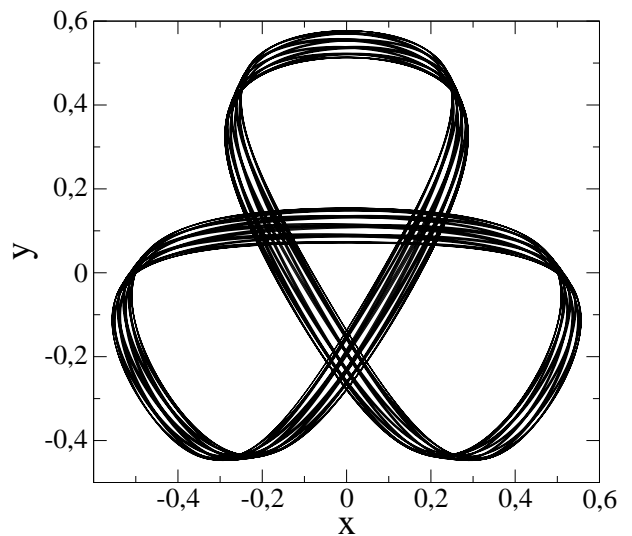}
\caption{\footnotesize{Trajetória quase-regular do sistema de Hénon-Heiles, 
com energia total igual a $1/6$.}}
\end{center}
\label{traj_16}
\end{figure}
\begin{figure}[ht!]
\begin{center}
\includegraphics[height=7cm,clip]{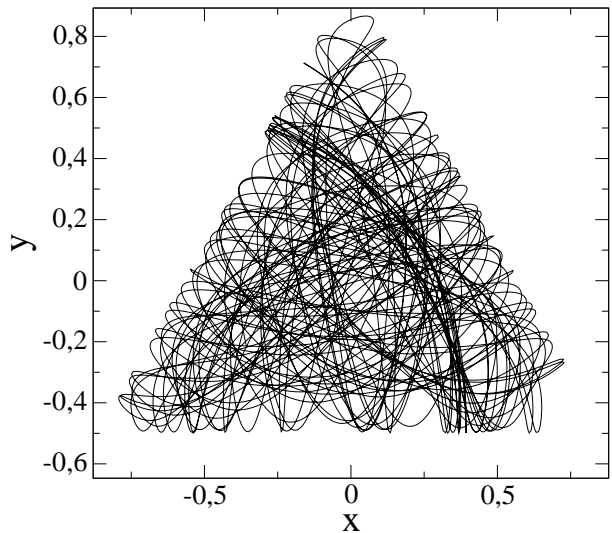}
\caption{\footnotesize{Trajetória caótica do sistema de Hénon-Heiles, 
com energia total igual a $1/6$.
}}
\end{center}
\label{poine_1610c}
\end{figure}
As duas trajetórias, figura (5) e (6), foram retiradas do sistema com energia 
$E=1/6$. Estas figuras mostram o espaço de configuração do sistema de 
Hénon-Heiles.
Observamos claramente, na figura (6), que a trajetória caótica não retornará 
a uma mesma posição $(x, y)$, enquanto a trajetória quase-regular, 
figura (5), se aproxima bastante da mesma região do plano $(x, y)$. 
Apesar de parecer que a trajetória quase-regular se fecha na figura (5), 
esta nunca retorna ao mesmo lugar.

Na figura (7), temos a seção de Poincaré para o sistema com energia $E=1/10$.
Nesta figura, pode-se observar a transição de fase do sistema, de integrável 
para misto, ou seja, o espaço de fases deixa de ter apenas ilhas fechadas para
apresentar a coexistência das fases regulares e caóticas. 
As trajetórias regulares e quase-regulares do sistema são representadas pelas 
``ilhas'' ou linhas fechadas, enquanto os pontos representam o ``mar'' caótico.
\begin{figure}[ht!]
\begin{center}
\includegraphics[height=7cm,clip]{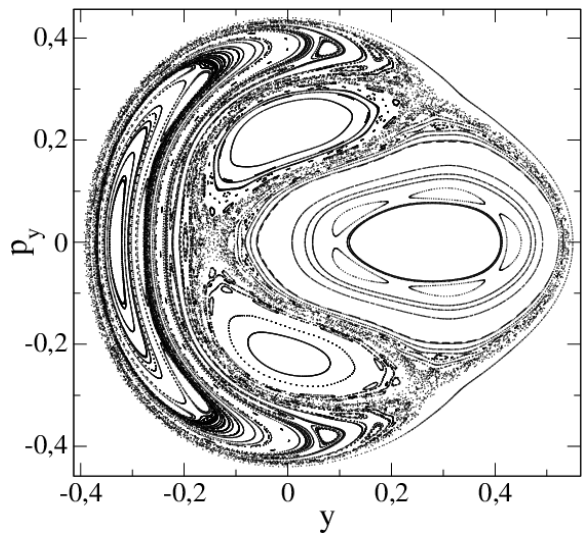}
\caption{\footnotesize{Seção de Poincaré para o sistema de Hénon-Heiles, com 
energia total igual a $1/10$ ou $E=0,1$, $10$ trajetórias e tempo de 
evolução de $4000$.}}
\end{center}
\label{poine_110}
\end{figure}
Visualiza-se, mais claramente, o mar caótico na figura (8), que é uma ampliação
da figura (7) para a região $-0,26<y<0,355$ e $-0.44<p_{y}<0.00$. Nesta figura 
pode-se observar alguns pontos elípticos (pontos de equilíbrio estável) nas 
coordenadas $(y, p_{y})= (0, -0,224)$, em $(0,085, -0,424)$, entre outros.

Nota-se que uma trajetória caótica (os pontos no retrato de fases) varrem
todo o espaço de fases, enquanto que a trajetória regular limita-se a 
determinadas regiões do espaço de forma compacta.
\begin{figure}[ht!]
\begin{center}
\includegraphics[height=7cm,clip]{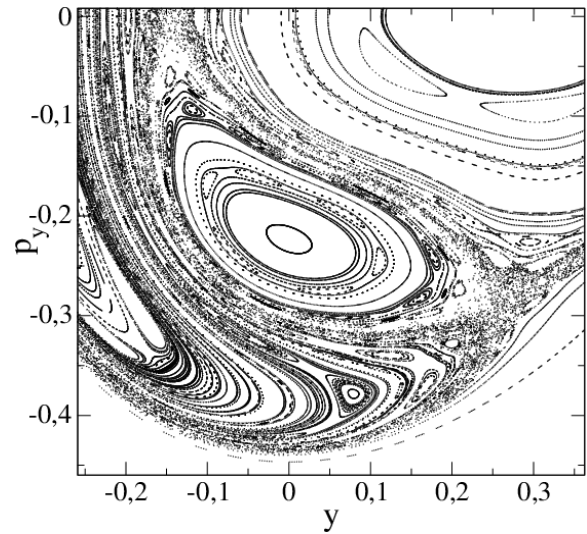}
\label{poine_1102}
\caption{\footnotesize{Ampliação da Seção de Poincaré para o sistema de 
Hénon-Heiles com energia total igual a $1/10$. Ampliação para ontervalo 
$-0,26<y<0,355$ e $-0.44<p_{y}<0.00$.}}
\end{center}
\end{figure}
\begin{figure}[ht!]
\begin{center}
\includegraphics[height=7cm,clip]{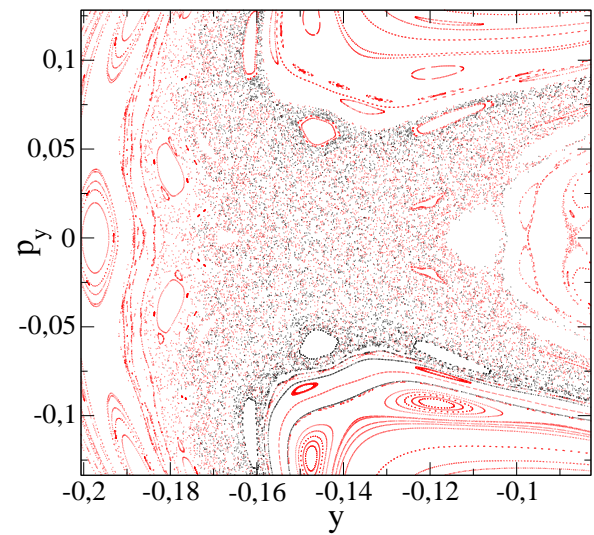}
\caption{\footnotesize{Seção de Poincaré para o sistema de Hénon-Heiles, com 
energia total igual a $1/10$, $50$ trajetórias e um tempo de evolução de 
$4000$. As ilhas de regularidade estão em vermelho e as trajetórias caóticas
estão em preto, bem como os stickiness. A figura também apresenta alguns pontos
em vermelho representando as trajetórias caóticas.}}
\end{center}
\label{poine_110s1}
\end{figure}
\begin{figure}[ht!]
\begin{center}
\includegraphics[height=7cm,clip]{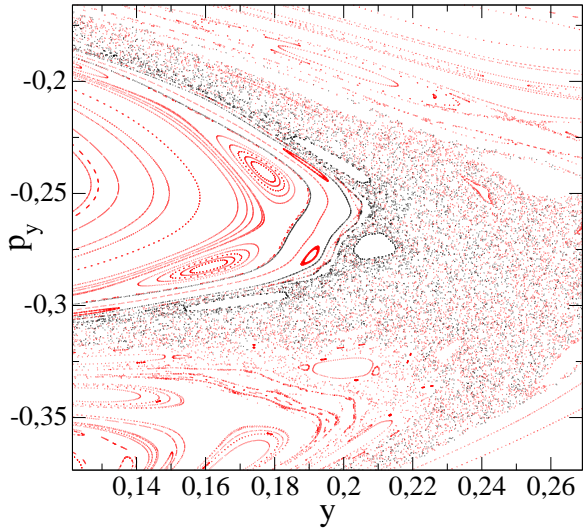}
\caption{\footnotesize{Seção de Poincaré para o sistema de Hénon-Heiles, com 
energia total igual a $1/10$, $50$ trajetórias e um tempo de evolução de 
$4000$. As ilhas de regularidade estão em vermelho e as trajetórias caóticas
estão em preto, bem como os stickiness.}}
\end{center}
\label{poine_110s2}
\end{figure}

A figura (9) apresenta uma ampliação da figura (7), com $50$ trajetórias, 
energia $E=1/10$ e tempo de evolução de $4000$, com coordenadas
$-0,21< y < -0,06$ e $-0.16< p_{y} < -0.16$. Esta e a próxima figura apresentam 
duas cores distintas no mesmo retrato de fases, preto e vermelho, para 
representar as trajetórias caóticas e regulares, respectivamente.
Pode-se observar que uma trajetória com coordenadas dentro do mar caótico,
percorre quase todo o espaço de fases, enquanto uma trajetória regular ou 
quase-regular limita-se as regiões onde estão as ilhas. Algumas trajetórias
caóticas também apresentam pontos em vermelho.

Dentro do mar caótico, na figura (9), 
observa-se uma densidade maior de pontos (região mais escura em preto) na
região $-0,16< y < -0,06$ e $\pm 0.10< p_{y} < \pm 0.05$, no entorno de ilhas.
Esses pontos são chamados de {\it Sticky}, ou {\it Grude} (tradução literal).
Esse nome vem do comportamento que esses pontos apresentam no espaço de fases.
Observa-se que eles ficam mais próximos às ilhas, representando um 
{\it grudamento} ({\it stickiness}) nas órbitas regulares. 
Para que o leitor se familiarize com o conceito e possa pesquisar 
posteriormente, utilisaremos os termos em Inglês.

O fenômeno de stickiness acontece quando algumas trajetórias se aproximam 
de ilhas de regularidades \cite{alt} (também chamadas de armadilhas dinâmicas 
\cite{zasl}) e passam um tempo considerável da sua evolução nessas regiões.
Isso faz com que o sistema fique mais próximo da regularidade nessas 
trajetórias. Um indicativo disto é o fato de que o expoente de Lyapunov (esta
ferramenta serve para quantificar o grau de caoticidade do sistema)
dessas trajetórias diminui \cite{her}.

Na figura (10) temos uma ampliação da figura (7), com as mesmas características 
da figura anterior, para as coordenadas $0,10< y < 0,27$ e 
$-0.32< p_{y} < -0.16$. Nesta figura são evoluidas $50$ trajetórias num tempo de
$4000$ quando o sistema tem energia $1/10$. Pode-se observar que existem regiões
caóticas e regulares na mesma seção de Poincaré e também o fenômeno de 
stickiness próximos às ilhas.
\begin{figure}[ht!]
\begin{center}
\includegraphics[height=7cm,clip]{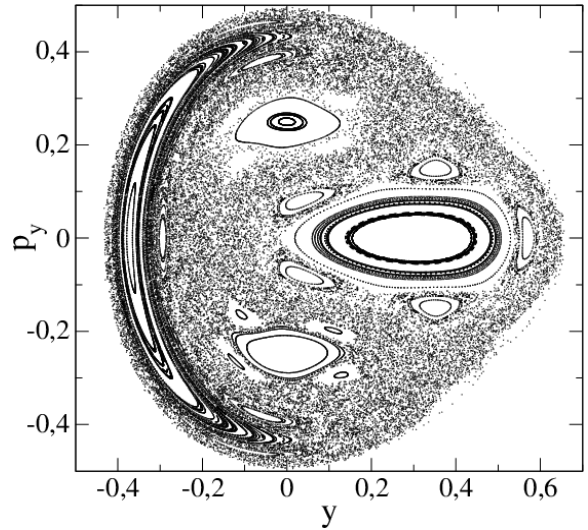}
\caption{\footnotesize{Seção de Poincaré para o sistema de Hénon-Heiles, com 
energia total igual a $1/8$, $110$ trajetórias e um tempo de evolução de 
$3000$.}}
\end{center}
\label{poine_18}
\end{figure}

Quando a energia do sistema cresce, mesmo que ele seja pequeno, 
o mar caótico aumenta, como pode ser visto na figura (11) (resultado
semelhante à referência \cite{chr}). 
Esta figura mostra a seção de Poincaré para o sistema de Hénon-Heiles 
para $110$ trajetórias, 
com condições iniciais aleatórias, tempo de evolução computacional de $3000$ e 
energia total $E=1/8=0,125$. A região caótica aumentou, mas ainda existem pontos
de estabilidade no sistema, representados através das ilhas.
Isso significa que, dependendo da condição inicial que tomamos nas coordenadas
$(x, y, p_{x}, p_{y})$, o sistema pode exibir um comportamento regular ou 
caótico. Uma partícula que se move sobre o plano descrito pela 
figura (3), pode se encontrar num ponto sobre este plano em que poderá 
permanecer em repouso, nas coordenadas $(x, y)$ e $p_{x}=0$, $p_{y}=0$, 
ou poderá mover-se pelo plano numa trajetória fechada, como na figura (5), 
esta ainda pode descrever uma trajetória irregular, sem retornar a mesma 
região inicial, percorrendo todo o triângulo como na figura (6).

\begin{figure}[ht!]
\begin{center}
\includegraphics[height=7cm,clip]{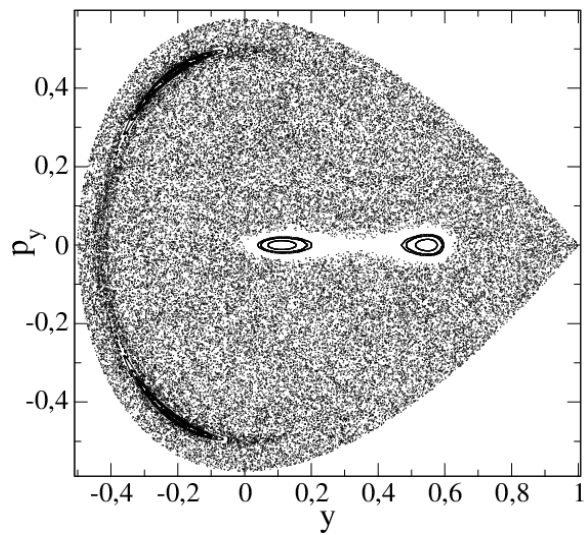}
\caption{\footnotesize{Seção de Poincaré para o sistema de Hénon-Heiles, com 
energia total igual a $1/6$, $10$ trajetórias e um tempo de evolução de 
$4000$.}}
\end{center}
\label{poine_16}
\end{figure}

Aumentando a energia do sistema para $E=1/6=0,1667$ e utilizando apenas 
$10$ condições
iniciais aleatórias para evolução temporal obtemos a seção de Poincaré da
figura (12). Esta figura apresenta uma expansão da região caótica e apenas
duas ilhas de regularidade menores, uma no centro do espaço de fases e outra
no canto esquerdo. A ilha central mostra que existem duas ilhas menores 
circundando dois pontos de equilíbrio \cite{chr}.
A ilha da esquerda está distribuída em $-0,5<p_{y}<0,5$, no entorno de um único 
ponto de equilíbrio centrado em $p_{y}=0$. 

O sistema tende a ficar totalmente caótico quando aumentamos mais a sua energia.
Quando o espaço de fases é tomado por pontos que representam o comportamento 
caótico o sistema é dito {\it ergódico}. 

\section{Conclusões}
\label{conclusoes}
Na última década investigações em sistemas Hamiltonianos tem apresentado 
grande interesse de pesquisadores da área de sistemas dinâmicos. 
A coexistência do comportamento regular e caótico no mesmo espaço de 
fases, o aparecimento de pontos elípticos e a presença de stickiness 
afetam o fenômeno de transporte em diferentes sistemas e podem explicar o 
comportamento de átomos confinados ou de dispositivos eletrônicos, como os 
chips feitos de semicondutores. 
Neste trabalho, analisamos o sistema de Hénon-Heiles em detalhes quanto ao seu
comportamento dinâmico. Descrevemos a Hamiltoniana do sistema e as equações
de movimento, através do formalismo Hamiltoniano. Verifica-se analiticamente
que o sistema é conservativo, com energia e momento angular constantes ao 
longo do tempo. 

Observa-se que o potencial de Hénon-Heiles apresenta formato triangular no 
plano e que esse sistema pode ser considerado um bilhar bidimensional.
Com diferentes condições iniciais e valores de energia observamos a transição 
das fases, de regulares à caóticas, no espaço de fases, quando a energia do 
sistema cresce. Para energia $E=1/24$ o sistema é integrável (regular). Quando
$E=1/10$ ou energias maiores o sistema apresenta comportamento caótico e 
regular para diferentes codições iniciais. 
No plano $(x, y)$ são analisadas dois tipos de trajetórias: as regulares 
(que formam ilhas no espaço de fases), e as caóticas (que formam pontos 
dispersos no espaço de fases). 
Observa-se o aparecimento dos grudes (ou sticky) na região caótica do 
espaço de fases próximos às ilhas de regularidade quando a energia do sistema 
aumenta. 

\section{Agradecimentos}
\label{agra}
O autor agradece a Fundação Araucária pelo suporte financeiro do projeto
20.029 do PPP/FA-14/2011.

\end{document}